\documentstyle{mn}

\title[Broken Homology and the Fundamental Plane]{Some Effects of Galaxy Structure and Dynamics on the Fundamental Plane}
\author[A.Graham and M.M.Colless]{Alister Graham$^{\star}$ and Matthew Colless\thanks{E-mail: ali@mso.anu.edu.au; colless@mso.anu.edu.au} \\
	Mount Stromlo and Siding Spring Observatories, The Australian National University, \\Private Bag, Weston Creek PO, ACT 2611, Australia.}
\date{ }

\begin{document}

\maketitle

\begin{abstract}

We examine the effects on the Fundamental Plane (FP) of structural departures 
from an $R^{1/4}$ galaxy light profile.  We also explore the use of spatial
(i.e.\ volumetric) as well as projected galaxy parameters.
We fit the Sersic $R^{1/n}$ law to the V-band light profiles 
of 26 E/S0 Virgo galaxies, where $n$ is a shape 
parameter that allows for structural differences amongst the profiles. 
The galaxy light profiles show a trend of systematic departures from a 
de Vaucouleurs $R^{1/4}$ law, in the sense that $n$ increases with 
increasing effective half-light radius $R_{\rm e}$.
This results in $R_{\rm e}$, and the associated mean surface brightness within
this radius, having systematic biases when constructed using an $R^{1/4}$ law.
Adjustments to the measured velocity dispersion are also made, based
upon the theoretical velocity dispersion profile shapes of the different 
$R^{1/n}$ light profiles, constructed assuming spherical symmetry and isotropic 
pressure support. 

We construct the FP for the case when structural homology is assumed 
(specifically, an $R^{1/4}$ law is fitted to all galaxies) and central 
velocity dispersions, $\sigma _{0}$, are used.  
The plane we obtain is 
$R_{\rm e}\propto \sigma _{0}^{1.33\pm0.10}\Sigma _{\rm e}^{-0.79\pm0.11}$,
where $\Sigma _{\rm e}$ is the mean surface brightness within the projected
effective radius $R_{\rm e}$. 
%(expressed in linear units rather than mag arcsec$^{-2}$).
This agrees with the FP obtained by others and departs from the virial theorem
expectation $R\propto \sigma ^{2}\Sigma ^{-1}$.
We find that allowing for broken structural homology through fitting $R^{1/n}$ 
profiles (with $n$ a free parameter), but still using 
central velocity dispersions, actually increases the 
departure of the observed FP from the virial plane --- the increase in effective 
radius with galaxy luminosity (and $n$) is over-balanced by an associated 
decrease in the mean surface brightness.

In examining the use of spatial quantities and allowing for the different
velocity dispersion profiles corresponding to the observed light profiles, 
we find
that use of the spatial velocity dispersion at the spatial half-light radius
decreased the departure of the observed FP from the virial plane.  
(Use of the spatial half-light radius and mean surface brightness term had no 
effect on the FP as they are constant multiples of their projected values).
Through use of the Jeans hydrodynamical equation, we convert
the projected central aperture velocity dispersion, $\sigma _{0}$,
into the infinite aperture velocity dispersion, $\sigma _{\rm tot,n}$ (which is 
equal to one-third of the virial velocity dispersion). 
Using both the $R^{1/n}$ fit and $\sigma _{\rm tot,n}$ we obtain
$R_{\rm e,n}\propto \sigma _{\rm tot,n}^{1.44\pm0.11}\Sigma _{\rm e,n}^{-0.93\pm0.08}$.  Making the fullest allowance for broken structural homology 
thus brings the observed FP closer to the virial plane, with the exponent
of the surface brightness term consistent with the virial expectation.

%XXX- point the way forward.
% We have not explored dynamical anisotropy, which may be a 
% remaining key to the observed departure from the virial plane seen in the 
% velocity dispersion exponent.
% Nor have we explored M/L variations in individual galaxies, or
% metallicity effects, etc.

\end{abstract}


\section{Introduction}

It has been known for some years that elliptical galaxies are well 
represented by a two dimensional surface in the space defined by their
observable parameters \cite{BaL83,Dj87a,Lyn8a}.  This two 
dimensional manifold, commonly shown in the logarithmic 3-space of radius 
($R_{\rm e}$), intensity ($I_{\rm e}$), and central velocity dispersion 
($\sigma _{0}$), has come to be known as the Fundamental Plane (FP) 
\cite{DaD87,7Sb87}. The existence of such a plane has implications for the 
formation and evolutionary processes of elliptical galaxies.   
Furthermore, the small scatter about the FP makes it a useful tool for
estimating distances to elliptical galaxies.  Acknowledging some degree of
scatter about the FP, either intrinsic to the galaxies and/or due to 
measurement
errors, the deviations from the FP have been interpreted as peculiar velocities
\cite{7Sa87,DaF90}.  For such methods of studying peculiar velocity 
fields and bulk flow motions, the exact FP used is a crucial factor.

Under the assumptions of kinematic and structural homology (i.e. identical
form and shape for the kinematic and density profiles) amongst galaxies,
the virial theorem predicts that galaxies will occupy a plane in log-space 
amongst the 
observables radius ($R_{\rm e}$), mass surface density ($\eta _{\rm e}$), and 
central velocity dispersion ($\sigma _{0}$) \cite{F7S87}.  
The transformation between this plane and the observed plane is given by 
the mass-to-light ratios, M/L, of the galaxies.  Furthermore, the tight 
constraints on the thickness of the observed plane and its 
negligible curvature, implies that M/L is a power law function 
of the observables. The mass-to-light ratio has important implications for 
galaxy formation \cite{F7S87,Dj87b} and so one would 
like to be able to test the assumptions of homology that go into the model.

The observed difference between the FP and the relation predicted from the
virial theorem is usually viewed as a systematic variation in the 
mass-to-light ratio along the FP, given by $M/L\sim L^{\alpha }$, where 
$\alpha \sim 0.25\pm 0.05$ \cite{FaJ76,F7S87,Dj87b}, and 
it has been shown to vary with bandpass (Djorgovski \& Santiago 1993; 
Pahre, Djorgovski \& de Carvalho 1995).
Many causes of such an effect have been explored and dismissed to varying
extents.  Possible stellar population differences along the FP 
\cite{Dj87b,DaS93,RaC93} seem unlikely due to the 
required fine-tuning of the IMF or $M/L$ required in order to maintain the 
thinness of the observed FP.  
A systematic variation in stellar age along the plane could produce the 
observed tilt, but would require a large conspiracy in the formation of
all galaxies of a given mass in order to be consistent with the observed 
small thickness of the FP.

To examine stellar population effects, Pahre et al.\ \shortcite{PDC95} tested 
the significance of the mass-metallicity relation \cite{Guz02} 
along the FP, by constructing a near-infrared FP.  Sampling the older 
distribution of stars, their K-band photometry is less sensitive than  
optical wave-bands to differences in galaxy metallicity which appear as
line-blanketing effects in the optical \cite{F7S87}.  The 
near-infrared FP obtained did not differ markedly from the optical FP, with 
the deviation found being attributed to the reduction in metallicity 
effects between galaxies.
This agrees with earlier work by Dressler et al.\ (1987b), 
Recillas-Cruz et al.\ \shortcite{Rec90}, and Djorgovski \& Santiago 
\shortcite{DaS93},
where the metallicity effect was shown to be insufficient to explain the entire
observed FP departure from the virial plane.  This is also shown to be the case
with current population synthesis models, where metallicity effects result in 
only a small tilt in the plane \cite{Wor94}.

This has led to investigations that test the assumptions of kinematic and
structural homology and measure the influence of such a breakdown upon
the FP (Ciotti, Lanzoni \& Renzini 1996; Hjorth \& Madsen 1995).  
Pahre et al.\ \shortcite{PDC95} also concluded that systematic 
departures from structural and dynamical homology amongst the galaxies may be 
responsible for the departure between the observed and the virial form of the 
FP.  

Capelato, de Carvalho \& Carlberg \shortcite{CCC95}, using computer model 
simulations of merged galaxies, claim 
that the slope of the FP is largely explained by broken homology in the velocity 
distribution.  However, Ciotti et al.\ \shortcite{CLR96} and Ciotti
 \& Lanzoni \shortcite{CaL96}, using spherical 
models with the velocity anisotropy described by the Osipkov-Merritt formula, 
find it is unlikely that orbital anisotropy plays a major role 
in producing the FP tilt.  
% However their investigation did not treat systems with any degree of 
%rotational support.  
Anisotropy in the velocity distribution has also been rejected by Djorgovski 
\& Santiago (1993) as the sole cause for the $M/L\sim L^{\alpha }$ relation.
Ciotti et al.\ \shortcite{CLR96} suggested that a systematic trend in the shape 
of the elliptical galaxy light profiles seems to be the best explanation. 

This possibility has been entertained before (Djorgovksi, de Carvalho, \& 
Han 1988; Djorgovski \& Santiago 1993), but differing profile shapes causing  
the tilt of the FP has not been explored with observational data.
Hjorth \& Madsen (1995) parameterised broken structural homology in their
models of elliptical galaxies based on the statistical mechanics of violent
relaxation \cite{HaM91}.  They found that for a given correlation between 
the galaxy structure and its luminosity, such non-homology is not inconsistent
with the observed FP having a constant $M/L$.  Unfortunately, their data set
did not enable them to draw firm conclusions.  There is, however, a growing 
wave of data supporting the existence of broken structural
homology, amongst dwarf galaxies \cite{Dav88,BaC91}, spiral bulges 
\cite{Cap87,Cap89,APB95,CJB96}, elliptical galaxies 
(Michard 1985; Caon, Capaccioli \& D'Onofrio 1993; Burkert 1993; 
Hjorth \& Madsen 1995), and brightest cluster galaxies \cite{Sch86,Gra96}. 

In this paper, using a set of observed surface brightness profiles 
and measured velocity dispersions, we explore the influence of broken 
structural and dynamical homology upon 
the FP.  Using the Sersic $R^{1/n}$ light profile \cite{Ser68}, 
the generalised form of the de Vaucouleurs $R^{1/4}$ law (de Vaucouleurs 
1948, 1953) with $n$ a free parameter, we can allow for 
structural differences between galaxies.  
The $R^{1/4}$ law has two parameters which act as physical scales, scaling 
the radius and the surface 
brightness and leaving the profile shape fixed.  The $R^{1/n}$ profile has the
additional parameter, n, which permits different shapes for the galaxy light
profile and therefore allows us to examine broken structural homology amongst 
galaxies.  
The Sersic law can take the form of both an exponential disk ($n$=1) and the 
de Vaucouleurs light profile ($n$=4).
It can also approximate a power-law and provides intermediate forms between
these common profile shapes by varying its shape parameter $n$.
Caon et al.\ \shortcite{CCDet} and Graham et al.\ \shortcite{Gra96}, 
and references therein, have shown 
that a range in galaxy profile shapes does exist, and more importantly that a 
systematic trend between profile shape and galaxy  size exists such that the 
larger galaxies have less curvature in their light profiles than the 
smaller galaxies.  In this paper we examine the effect of this on the FP.
We also explore the use of volumetric galaxy parameters as well as using
the standard projected quantities.  It is possible to obtain the three 
dimensional structural quantities by deprojecting the $R^{1/n}$ light 
profiles 
(Ciotti 1991).  The volumetric kinematical quantity is derived from 
application of the Jeans hydrodynamical equation to the $R^{1/n}$ model,
calibrated by the projected central velocity dispersion.

In section 2 we describe our sample of elliptical galaxies and the derivation 
of their structural parameters.  An overview of the parameters used in the 
construction of the FP and their relation to the ideal virial plane parameters 
is given in section 3.
We use multivariate statistics to describe and construct the FP
in section 4.  Our results are discussed in section 5, and our conclusions
presented in section 6.

\section{Data}
\subsection{Surface Brightness Profiles}

For the purposes of this investigation we have chosen the small but very 
high-quality data set of Bower, Lucey \& Ellis \shortcite{BLE92}.
The aperture magnitude profiles of 26 E/S0 Virgo cluster galaxies
were selected, due to the quality of this CCD 
data (with an rms internal scatter of 0.015 mag) and as published dynamical 
data is also available for these galaxies (Dressler 1984, 1987; McElroy 1995).

We took the tabulated V band profiles, and corrected them for the galactic
absorption and redshift terms as follows.  We applied the K-correction as 
detailed by Bower et al.\ \shortcite{BLE92}, being $\Delta V$=0.005 mag for 
Virgo.  Galactic extinctions in B-band ($A_{B}$), taken from Burstein \& 
Heiles \shortcite{BaH84} and Burstein et al.\ \shortcite{B7S87}, were 
applied to our data such that $A_{V}=0.75A_{B}$ \cite{San73}.
For those galaxies not listed, we used the NASA/IPAC Extragalactic DataBase 
(NED) to find their extinction.
We also applied the corrective term for redshift dimming, being -0.013 mag for
Virgo at a redshift of 0.003.  So as to remove the data most affected by 
seeing, we truncated the inner profile, excluding data within 3 seeing FWHM.  
Each galaxy profile consisted of 14 data points before allowances were
made for seeing, which resulted in the removal of between 2 to 5 inner data 
points.
The outer data point is given by the largest aperture fitted by Bower
et al.\ \shortcite{BLE92}, having a diameter of 79 arcsec.

\subsection{Galaxy Parameters}

We have used the velocity dispersion data in Table 11 of Bower et al.\ 
\shortcite{BLE92}, which comes from
Dressler (1984, 1987).
%\shortcite{Dre84}, \shortcite{Dre87}.
We note that an effective 
aperture of 16$\times$16 arcsec was used to make these
measurements (Dressler 1984).  Four of the galaxies were not measured by 
Dressler (marked in Table 2) and so 
we have used the mean central velocity dispersion given by McElroy \shortcite{McE95},
correcting it to the standard used by Dressler.  From Table 2 of McElroy \shortcite{McE95},
this implies dividing the central velocity dispersion value of McElroy 
by 1.02 to obtain the value which would have been measured using Dressler's
system.  
% 3 of these 4 galaxies have velocity dispersion measures less than 
% 100 km s$^{-1}$.  Such low values have meant the exclusion of such galaxies from FP
% analysis in the past, due to the large scatter in the plane at this low end.
% However, we opt to keep these galaxies as we plan to convert these 
% measurements to a new standard that has not been done before (see the 
% following section).  That is maybe there is the larger scatter at the
% faint end of the FP because the velocity dispersion term has not been 
% treated properly in the past.  [check by looking at the planes I construct].

We obtained the effective radius, $R_{\rm e}$, and mean surface 
brightness, $\Sigma _{\rm e}$, within $R_{\rm e}$ (expressed in linear units rather 
than mag arcsec$^{-2}$) by fitting an $R^{1/n}$ model profile 
\cite{Ser68} to the corrected luminosity profiles.  The form of this 
profile is
\begin{equation}
I(R)=I_{\rm e}\exp\left\{ -b\left[\left( \frac{R}{R_{\rm e}}\right) ^{1/n} -1\right]\right\}, \label{sereq}
\end{equation}
where $I_{\rm e}$ is the intensity at the radius $R_{\rm e}$.  The parameter $b$ is
chosen so that $R_{\rm e}$ is the projected radius enclosing half the 
total light from the galaxy and is well approximated by $b=1.9992n-0.3271$ 
\cite{Cap89}.  

The aperture magnitude, within radius $R$, is obtained from the projected
luminosity within a circular aperture of this radius.  The luminosity
within a radius $R$, is then calculated from $2\pi\int _{0}^{R}I(R)R\,dR$.  
Using the change
of variable $x=b(R/R_{\rm e})^{1/n}$, it can be shown that
\begin{equation}
L(R)=2\pi n\frac{e^{b}}{b^{2n}}\gamma (2n,x)I_{\rm e}R_{\rm e}^{2}, \label{ltot}
\end{equation}
where $\gamma $ is the incomplete gamma function.  The galaxy light profiles
of Bower et al.\ \shortcite{BLE92} are in the form of aperture magnitudes, 
and we used a 
standard non-linear least squares fit (i.e. minimisation of the ordinate 
to define the best fitting line) to obtain the model parameters
$R_{\rm e}$, $\mu _{\rm e} (=-2.5\log I_{\rm e}$), and $n$, from the equation
\begin{equation}
m(r)=C-2.5\log \gamma (2n,x),
\end{equation}
where $C=-2.5\log[I_{\rm e}R_{\rm e}^{2}2\pi n e^{b}/b^{2n}]$.
The residuals of the aperture magnitude data points about the best-fitting 
$R^{1/n}$ profiles are shown in Figure~\ref{curve}.  There were multiple observations 
of individual galaxies, and we show the residual profiles for each set of
data for each galaxy.

The model parameter $R_{\rm e}$ (arcsec) was converted to kpc using a Virgo 
redshift of 0.003, and $H_{0}=75$ km s$^{-1}$ Mpc$^{-1}$.
As $R_{\rm e}$ is sometimes larger than our observed outer radius data point 
of 39.5 arcsec, 
one cannot always simply integrate the profile data out to $R_{\rm e}$ in order 
to get $\Sigma _{\rm e}$.  An alternative approach was developed based upon the 
surface brightness profile from the $R^{1/n}$ model (see Appendix A).
Such extrapolation is valid given the assumption that the $R^{1/n}$ model
is suitable beyond our observed outer radial point and out to the model 
half-light radius $R_{\rm e}$.  For Virgo, 39.5 arcsec corresponds
to 2.28 kpc, and roughly 1/3 of our sample required an extrapolation.
We averaged the model parameters when multiple observations existed, 
obtaining a single mean value to represent each galaxy.
Agreement between different observations of similar galaxies is good.   
This can be seen in Table 1, or more easily in Figure~\ref{error}, where the 
1$\sigma \, \chi ^{2}$ confidence intervals of the fitted model parameters 
overlap each other for multiple images of the same galaxy.

We have plotted the logarithm of the average shape parameter $n$, obtained from the models fitted to 
the galaxy profiles, against the logarithm of the average spatial half-light radius in 
Figure~\ref{enrad}.  As seen in the following section, the spatial half-light 
radius is approximately 1.34$R_{\rm e}$ for all values of $n$.  That is not to
say the spatial half-light radius is independent of $n$, as it is 
$n$ that determines the profile shape and thus the value of $R_{\rm e}$
and the spatial half-light radius.
Also shown in the $n-\log R_{\rm e}$ plane are the $1\sigma $ 
confidence regions for the fitted model parameters of every image.  These 
are based upon the projected $\Delta \chi ^{2}=2.30$ contour around 
the best fitting solution ($n, R_{\rm e}, \mu _{\rm e}$), after normalising the 
reduced $\chi ^{2}$ to 1 at the fitted minimum (Figure~\ref{error}).
% The reduced
% $\chi ^{2}$ value is of the order $10-10^{2}$.  This result was found in
% \cite{Gra96}, and reflects the fact that the galaxy profile has wiggles
% in it at a level greater than can be accommodated for by our model fits.
% These may be caused by dust features, shells, or some other physical
% presence or influence that produces such ripples.  
% This situation has allowed us to apply our $\chi ^{2}$ analysis, 
% assuming normally distributed errors. (given systematic errors don't
% dominate ??? ).
Saglia et al.\ \shortcite{Sag96} point out that under certain conditions, such $\chi ^{2}$ 
contours can underestimate the true errors. 
This effect becomes increasingly apparent as the actual galaxy light profile 
departs further and further from the 
assumed model profile (an $R^{1/4}$ law in their analysis of 
this effect).  In fitting an $R^{1/n}$ model, we explicitly allow for such
an effect (to the extent that the $R^{1/n}$ model can correctly fit the
data) and hence reduce the influence of this effect on our estimated errors.
Saglia et al.\ \shortcite{Sag96} note that the discrepancy is 
mainly due to large extrapolations of the fitted profile models in obtaining
quantities like total galaxy magnitude.  We do not compute the total galaxy 
magnitude, and 
consequently our model parameters are largely free from such extrapolations.
Although $R_{\rm e,n}$ is the half-light radius it does {\bf not} require the total
magnitude for its derivation, which comes directly from the luminosity
profile fit.
A few galaxies did require some extrapolation, but only as far as their
model half-light radii.  We note that these model half-light radii may
not be the true galaxy half-light radii, as we don't have detailed extended
photometry and the $R^{1/n}$ model may not be appropriate for the faint outer
parts of the galaxy, where galaxy envelopes or halos may result in a 
departure from the inner profile shape.  
The model radii obtained {\em are} however representative of the brighter 
inner portions of the galaxy profile and are useful for our analysis.
Whilst there is a coupling between the galaxy model parameters (although
we note that a couple of the larger contours in Figure~\ref{error} have the 
effect of dominating this picture), such coupling cannot fully 
explain the observed trend.
The value of $n$ clearly increases with effective radius.

Bower et al.\ \shortcite{BLE92} made multiple observations of the 26 galaxies in the data 
sample.  Table 1 lists the model parameters for 
every observation.  The average values used in the construction of the FP
are listed in Table 2, along with the observed central velocity dispersions.
The values in Table 2 have been transformed from projected 
quantities into spatial quantities, as detailed in the following section.

\subsection{Possible sky-subtraction errors}

We explore the possibility that sky-subtraction errors may be leading
to the observed departures from an $R^{1/4}$ profile.  
Even with all of the care that the Bower et al.\ \shortcite{BLE92} data was put together, the
use of small area detectors can leave residual sky-subtraction errors.
Perhaps the larger galaxies, with apparently greater values of $n$, are 
a product of systematically larger sky-subtraction errors which mimic 
departures from an $R^{1/4}$ law.
% -- as the CCD frames may not contain edges or corners completely free from galaxy light.

To investigate this, we fitted the galaxy profile data with the model
\begin{equation}
I(R)=I_{\rm e}\exp\left\{ -7.669\left[\left( \frac{R}{R_{\rm e}}\right) ^{1/4} -1\right]\right\} + I_{sky}, \label{skyeq}
\end{equation}
which is the standard $R^{1/4}$ law plus an intensity term for the possible
sky-subtraction residual (which can be either positive or negative in this
equation).  Sky-subtraction is a common concern, and an incorrect value for
the sky brightness will result in changes to the derived galaxy magnitude 
profile.  The addition of an extra parameter in the profile fitting, namely
$I_{sky}$, will of course produce more accurate fits as governed by the
$\chi ^{2}$ statistic.  The question is whether or not this extra
parameter (actually a correction) is justifiable.

Plotted in Figure~\ref{chise} is the ratio of the $\chi ^{2}$ fit from
the $R^{1/4}$ + sky-correction model to the $\chi ^{2}$ fit from the
simple $R^{1/4}$ model.  Also shown is the ratio of the $\chi ^{2}$ fit from
the $R^{1/n}$ model to that from the simple $R^{1/4}$ model.
Both of the additional models, containing a third parameter,
clearly show an improvement in fit when one deviates from an $n$=4 profile.
The question to be addressed here is whether use of the sky-corrected 
$R^{1/4}$ model can be substantiated.
We turn our attention to the size of the sky-subtraction correction
that these models imply.  

The photometry comparison with previous authors work, which
Bower et al.\ \shortcite{BLE92} performed, puts a limit on the size of possible errors.
Even if we attribute the entire disagreement between observers work to
sky-subtraction errors, we still have quite stringent limits.  
Aaronson, Persson \& Frogel \shortcite{APF81}, using large photographic plates rather than small area CCD's,
obtained V-band data for 22 Virgo galaxies in common with the sample of Bower 
et al.\ \shortcite{BLE92}.
Rejecting 2 out lying points, Bower et al.\ \shortcite{BLE92} found a scatter 
of only 0.035 mag in the comparison photometry.
A similar scatter of 0.036 mag was found with the comparison of
Michard \shortcite{Mic82} (also with 20 common galaxies).
Even if we attribute all of this scatter to sky-subtraction errors by
Bower et al.\ \shortcite{BLE92}, we still have a very tight constraint on the 
allowed sky-subtraction errors in our profile models.  Figure~\ref{ratts} 
shows 
that the majority of sky-subtraction errors are notably larger than permitted
 by our quite liberal error allowance.  So while the addition of 
a sky-subtraction correction produces improved fits to the data,
it is not justified.  

The introduction of this 
additional parameter has the success it does because it is mimicking,
to some degree, the effect of some other real physical process or
situation.  Whilst an $R^{1/4}$ law is being fitted in these models, it
is being done coincidentally with a term that effectively re-shapes the
profile to obtain an $R^{1/4}$ profile.  In this fashion, the model 
incorporating a sky-correction term is able to allow for real differences 
in galaxy profile shapes, as better represented by the $R^{1/n}$ law
(Caon et al.\ 1993; Graham et al.\ 1996)
In Figure~\ref{ratio}, we show that the 
$R^{1/4}$ + sky-correction model is only superior to the $R^{1/n}$ model
for unacceptably large under-estimations of the sky background level.
Also, 13 of the points  in Figure~\ref{ratio} for which the $R^{1/4}$ + 
sky-correction model is superior to the $R^{1/n}$ model belong to just
one galaxy (NGC 4552), which was observed 14 times; 6 of these points occupy
the cluster of 8 points that favour the $R^{1/4}$ +
sky-correction model.  Removal of NGC 4552 eliminates the impression that
use of an $R^{1/4}$ + sky-correction model is generally better than the
$R^{1/n}$ model when one is dealing with the larger galaxies ($n$$>$$4$).

In further support of our conclusion that sky-subtraction errors are not
the effect responsible for the range in profile shapes seen, Capelato et al.\
 (1995) also found a correlation between $n$ and $R_{\rm e}$.  The $R^{1/n}$
models they fitted to their computer simulated elliptical galaxy merger
remnants, have no uncertainty as to sky levels, thus 
eliminating this possibility.  There is however another possibility as to 
the cause of the range in our profile
shapes.  S0 galaxies are known to possess a disk as-well as their central 
bulge.  If our sample of galaxies contained significant disks and a range
of bulge-to-disk ratios, then this might explain the apparent broken 
structural homology \cite{Sag96}.  
Whilst this could be so, an argument against such an
origin is that the galaxy models of Capelato et al.\ (1995) were disk 
free, as was the work with Brightest Cluster Galaxies by Graham et al.\ 
\shortcite{Gra96}, where a large range in profile shapes was still evident.  
Also, given the observed trend with $n$ and $R_{\rm e}$, one would have to claim
that the disk-to-bulge ratio increased for the bigger galaxies, which it does 
not \cite{APB95}.
%  and so the contaminating disk explanation for our E/S0 
% galaxy sample seems a bit improbable.
% Although it might be possible that both effects (bulge-disk composition 
% with $R^{1/n}$ interior) could be present in our galaxy sample, the use of an 
% $R^{1/n}$ model to represent the observed departures from structural homology 
% will serve our purposes.

\section{Theory}
%  Numerous processes can be thought of that could disrupt the existence of the FP.

\subsection{Overview}

A good summary of the relation between the observed and the virial planes
can be found in Djorgovksi, de Carvalho \& Han \shortcite{DDH88}, and we 
adopt their method of
analysis in what follows.  From the virial theorem, the connection between 
the potential and kinetic energy can be written as
\begin{equation}
\frac{GM}{<R>}=k_{\rm e}\frac{<V^{2}>}{2},\label{viril}
\end{equation}
with $M$ and $<$$R$$>$ being a measure of the galaxy mass and radius, respectively,
 such that $GM/$$<$$R$$>$ is the galaxy potential energy 
and $<$$V^{2}$$>$$/2$ is the mean kinetic energy per unit mass for the galaxy.  
$k_{\rm e}$ is a 
virialisation constant reflecting the degree to which the system is virialised. 
%(If $M, <$$R$$>$ and $<$$V^{2}$$>$ are the virial quantities, $k_{\rm e}>1$ for a 
%bound system and equal to 2 for a virialised system) 
Galaxy mergers and the process of violent relaxation redistribute the orbital
energy of the stars.  Galaxies may exist in a 
quasi-static equilibrium, with varying degrees of energy dissipation or in
varying stages of relaxation, and hence varying values of $k_{\rm e}$ may exist
before a galaxy settles down into virial equilibrium ($k_{\rm e}$$=$$2$)
\cite{DDH88}.
In addition to this, the $k_{\rm e}$ term also allows for possible neglect of
rotational energy: in our use of central velocity dispersions to measure the 
kinetic term, $<$$V^{2}$$>$, we have ignored possible contributions from the 
rotational energy of each galaxy.

A choice now arises regarding which observables one uses to fit for the
quantities $M$, $<$$R$$>$ and $<$$V^{2}$$>$.

\subsection{Structure}\label{3two}

For elliptical galaxies, an observed radius $R$ (such as $R_{\rm e}$) 
is typically used to approximate $<$$R$$>$ such that 
\begin{equation}
R=k_{R}<R>.  
\end{equation}
The parameter $k_{R}$ allows for the possibility that our
observed projected radius R may not be equal to the physical radius we wish
to use.  $k_{R}$ depends on the mass-density 
structure of the galaxy, allowing for variations in the distribution of
matter from one galaxy to the next.  It is possible that different galaxies 
formed with a different mass-density structure. It is also likely that 
evolutionary processes such as galaxy merging and tidal interactions have 
further modified the $k_{R}$ term.  By addressing the breakdown of structural
homology between galaxies, we can explore the variations in the $k_{R}$ term.
These variations, if ignored, introduce systematic errors into the 
parameterisation of elliptical galaxies, and hence into the construction of
the FP.

The central velocity dispersion, $\sigma _{0}$, is used to represent the 
kinetic energy term such that 
\begin{equation}
\sigma _{0}^{2}=k_{V}<V^{2}>,
\end{equation}
with $k_{V}$ reflecting the dynamical structure within the galaxy.  
Like the $k_{R}$ term, the $k_{V}$ term reflects the formative and
various evolutionary processes that have shaped the galaxy.
It incorporates the anisotropies of the velocity dispersion tensor, including
the differing degrees of random and circular motion in the galaxy, and
differing velocity dispersion profiles.

Lastly, we have 
\begin{equation}
(M/L)\Sigma R^{2}=\frac{1}{k_{L}}M,\label{mass} 
\end{equation}
where (M/L) is the global galaxy mass-to-light ratio, and $k_{L}$ incorporates 
the luminosity structure within the galaxy.  $\Sigma $ is an intensity term, 
obtained from the fitted light profile.

Substitution into equation~\ref{viril} gives
\begin{equation}
R=K_{SR}\sigma _{0}^{2}\Sigma ^{-1}(M/L)^{-1},\label{kvirl}
\end{equation}
with
\begin{equation}
K_{SR}=\frac{k_{\rm e}}{2Gk_{R}k_{L}k_{V}}\label{kkkk}
\end{equation}
being the combined structural term appropriate for this expression, 
as given by Djorgovski et al.\ \shortcite{DDH88}.
Without knowledge of the $K_{SR}$ term, it has, in the past, been assumed a 
constant, and the observed departures of the FP from the relation 
$R\propto \sigma _{0}^{2}\Sigma ^{-1}$ have been attributed to (M/L), 
perhaps with
an additional comment as to uncertainties about assumed structural and 
dynamical homology.  This paper advances the work of Djorgovski et al.\ 
\shortcite{DDH88}, where
they proceeded under the assumption of homology (i.e. that all the $k$'s are
constant) - we do not.
Through fitting an $R^{1/n}$ profile which specifically allows for
structural differences, there is no assumption of structural homology.
We point out that whilst the $R^{1/n}$ profile covers an extensive range 
of profile shapes, 
it does not cover every possible profile shape as that would require an 
endless number of parameters in the fitted profile model.  Thus, we only
explore broken structural homology as far as the $R^{1/n}$ model allows,
which turns out to be quite sufficient.
Allowing for variability in the profile shapes, we can now better address 
the $k_{R}$ and $k_{L}$ terms, and 
also have a better representation for the $k_{V}$ term.

%Consider a particle (e.g. star) in motion throughout a galaxy at 
%an orbital radius $r_{1/2}$ equal to the spatial (i.e. three-dimensional) 
%half-mass radius,
%which is equivalent to the spatial half-light radius for a constant M/L
%(n.b. this is not the projected half-light radius on the sky).  We then have 
%$<R>=R_{1/2}$, $M=M_{tot}/2$, and $<V^{2}>$ being given by the spatial 
%velocity dispersion at $r_{1/2}$, for equation~\ref{viril}.

We explore the use of spatial (i.e.\ three-dimensional) quantities to represent 
$<$$R$$>$$, M$, and $<$$V^{2}$$>$.  We take the spatial half-mass radius, 
$r_{1/2}$,
which is equivalent to the spatial half-light radius for a constant M/L,
to represent $<$$R$$>$.
We compute the total galaxy luminosity, and use the assumption of constant
M/L to obtain half the total galaxy mass which is used to represent $M$. 
The spatial velocity dispersion at $r_{1/2}$ is acquired using the Jeans
equation (see Section 3.2) and is used as our estimate of $<$$V^{2}$$>$.
These quantities are not directly observable, but they can be computed from
other observable quantities given a knowledge of the k-terms.

Ciotti \shortcite{Cio91} related the luminosity form of the Sersic law to its
underlying mass-density structural form, by solving the Abel integral equation
\cite{BaT87} which relates the projected/deprojected expressions.  
He also computed the spatial half-mass 
radius $r_{1/2}$ for differing exponents $n$, and found that this radius, 
divided by the projected effective radius $R_{\rm e}$, is practically a constant 
value (ranging from 1.339 to 1.355 for $n$=2 to 10). 
 We note a similar figure was also found for the Jaffe \shortcite{Jaffe} models (1.311) 
and the Hernquist \shortcite{Her90} models (1.33).  Thus, the spatial half-light radius 
(in units of $R_{\rm e}$) is largely independent of the variant forms of the 
mass-density distributions used to describe the galaxy and so $k_{R}$ can 
be taken to be a constant with 
\begin{equation}
R=R_{\rm e,n}\approx (1/1.34)<R>=0.75 r_{1/2}.
\end{equation}
This is not to say that $r_{1/2}$ is independent of $n$ and therefore does 
not depend on the galaxy profile shape. What it does say is that $r_{1/2}$ 
is dependent on the
projected half-light radius, which is very much dependent on the profile 
shape and hence the value of $n$.

The structural term $k_{L}$ is not as simple, but it can be computed for 
different values of $n$.  Using $M=M_{tot}/2$ and (M/L)=($M_{tot}/L_{tot}$)
in equation~\ref{mass}, upon substitution of $L_{tot}$ from 
equation~\ref{ltot} (where the incomplete gamma function, $\gamma $, becomes
the gamma function $\Gamma $), we have 
\begin{equation}
k_{L}=\pi n\Gamma (2n)e^{b}/b^{2n}.
\end{equation}
As $b$ is well approximated by a simple linear function of n (1$<$$n$$<$10), 
we have an expression for the luminosity 
structural term in terms of the shape parameter $n$.
Allowances for different galaxy luminosity structures are made to the mean
intensity term $\Sigma $, which is fed into the FP as the surface brightness 
parameter.   This is achieved by using $\mu _{\rm e}-2.5\log (k_{L})$, rather 
than $\mu _{\rm e}$ (the surface brightness at $R_{\rm e}$) which does not fully 
represent the luminosity structure
(as detailed in equation~\ref{mass}).  If all galaxies did obey an $R^{1/4}$ 
law, then the difference between these terms would just be a constant and 
use of one term over another would not effect the tilt of the FP.

The mean surface brightness, $\langle \mu \rangle _{\rm e}$, within the effective 
radius, $R_{\rm e}$, is related to the intensity by $-2.5\log 
\langle I \rangle _{\rm e}$, where $\langle I \rangle _{\rm e}=L_{tot}/2\pi R_{\rm e}^{2}$ from 
Appendix A.  In Appendix A, we also see that $\langle \mu \rangle _{\rm e}-[\mu _{\rm e}
-2.5\log (k_{L})]=2.5\log(\pi )$, independent of the value of $n$.  
Thus $\langle \mu \rangle _{\rm e}$ can be
used in the construction of the FP, instead of $\mu _{\rm e}-2.5\log (k_{L})$.
The intercept of the FP will be effected but not the slope of the plane.

Thus, to the extent that the $R^{1/n}$ model represents the observed 
variations in structural
form, we are able to address two of the components that go into making the
$K_{SR}$ term which determines differences between the observed and the virial 
plane.  The $R^{1/n}$ model parameters, $R_{\rm e}$ and $\langle\mu\rangle_{\rm e}$ 
implicitly allow for these variations in profile shape which is explicitly 
represented by the third model parameter $n$.

\subsection{Dynamics}

As we do not confront the issue of orbital anisotropy, we cannot fully 
address the $k_{V}$ term, although we do partially deal with it by tackling
the problem of differing velocity dispersion profile shapes due to differing
luminosity profiles, as represented by the $R^{1/n}$ model.
Dynamical homology implies that each galaxy has the same fall off in 
velocity dispersion away from its center. 
It is therefore obvious that broken structural homology necessitates broken 
dynamical homology.
Working with the assumptions of spherical symmetry and isotropic pressure
support, Ciotti
\shortcite{Cio91} has shown how both the spatial velocity dispersion profile 
and the projected line of sight velocity dispersion profile change
with radius for a series of $R^{1/n}$ models with a range of shape parameters 
from 2 to 10.  
We do not deal with possible dynamical anisotropies within
each galaxy, as this requires detailed spectral mapping and/or assumptions
about the ratio of the radial to azimuthal velocity dispersion (Ciotti, 
Lanzoni \& Renzini 1996).
%Varying degrees of dynamical anisotropy may also be present amongst the 
%galaxy population and this will also increase the level of dynamical 
%non-homology

There is of course the additional concern of an aperture effect.  Even if all
galaxies did have the same velocity dispersion profiles, as one samples 
galaxies that are further away, the fixed aperture size on the sky will sample
a larger portion of the galaxy.  This has been addressed by Dressler (1984)
and Davies et al.\ \shortcite{Dav87}, and is implicitly taken into account in 
what follows.  However, the problem is more complicated than this.
What one would actually like to do, is not be consistent in using
the same physical aperture size on each galaxy, but to sample each galaxy 
with the same
ratio of aperture size to half-light radius\footnote{This correct approach would imply a different conversion of the central velocity dispersion in McElroy 
\shortcite{McE95} to the standard of Dressler \cite{Dre84}.} \cite{Weg96}.  
That is, to always measure the velocity dispersion within say, 1$R_{\rm e}$ 
or 0.2$R_{\rm e}$ or some fixed ratio of $R_{\rm e}$.
It is clear that use of core velocity dispersion measurements is inadequate in
representing a galaxies overall kinematic state.

Extending the work of Ciotti \shortcite{Cio91} we compute the projected aperture 
velocity dispersion as a function of the ratio of the aperture size to the 
projected half-light radius, as shown in 
Figure~\ref{vel} (The details are left for Appendix B).
From these models, the ratio of the 
projected aperture velocity dispersion (of aperture size equal to that used
to obtain the central velocity dispersion at the telescope) to the
spatial velocity dispersion at $r_{1/2}$, can be used to correct the 
measured central velocity dispersion to the spatial half-mass value.  
This ratio is shown in Figure~\ref{perc}.
Depending on the aperture size used (actually the ratio of the aperture
size to the projected half-light radius), use of the measured central
velocity dispersion by itself may result in departures of a factor of 2
or 3 from the spatial half-mass velocity dispersion measure.

It is possible to estimate the galaxy masses, using our measures of galaxy
size and velocity dispersion.  Once we have fitted the galaxy luminosity
profile with an $R^{1/n}$ law, the only unknown in equation~\ref{maseq} is
($M/L$), where we have used our measured central velocity dispersion for 
$\sigma ^{2}_{\rm ap,n}(R_{\rm ap})$ and the value predicted by theory, shown in 
Figure~\ref{vel}, for $\sigma ^{2}_{\rm ap,n}(\eta _{\rm ap})$.  
Substituting in the total galaxy luminosity for $L$, from 
equation~\ref{ltot}, we can then solve for the total galaxy mass $M_{tot}$.
Figure~\ref{massn} shows the total galaxy mass obtained in this way, plotted
against the shape parameter $n$.  
% The outlying points are NGC4486 (low n, high
% mass), NGC4377 (high n, low mass), NGC4382 (high n, high mass).
%  (we have used the terms low and high only to aid the reader with the 
% identification of the data points, we do not imply that these galaxies 
% have unreasonably low or high values of $n$ or $M_{tot}$.)

We also use the model velocity dispersion profiles and the observed central 
velocity dispersion to compute the aperture velocity dispersion extended to
infinity, $\sigma _{tot,n}$.  Ciotti (1994) shows this quantity to be equal to
one third of the virial velocity dispersion (in the case of spherical systems)
and independent of any possible orbital anisotropy.

Finally, we remind the reader that this paper does not treat the fourth 
parameter $k_{\rm e}$, in equation~\ref{kkkk}, which allows for different degrees
of virialisation and the neglect of rotation.

\section{Fundamental Plane construction}

We construct four planes.  The first one is the standard plane, obtained
by fitting $R^{1/4}$ profiles, and uses the projected effective 
radius $R_{\rm e}$, the mean surface brightness within this radius, $\Sigma _{\rm e}$,
and the measured central aperture velocity dispersion $\sigma _{0}$.  This 
plane is used as our reference plane, to see how the planes produced by
alternate choices of variables change.
% We note here, that use of the projected half-light radius and the mean surface
% brightness within this radius produces the same plane as use of the 
% spatial half-mass radius and central surface brightness or surface brightness
% at $R_{\rm e}$ when an $R^{1/4}$ is applied to the galaxy luminosity profile.
% This is because ... 
The second plane that we construct allows 
for differences in galaxy structure.  We fit an $R^{1/n}$ profile, and use the
spatial half-light radius, the associated mean surface brightness within this
radius, and the central aperture velocity
dispersion.  Such a plane will tell us the extent to which use of an $R^{1/4}$
law for all galaxies is responsible for the tilt of the FP.  
The third plane treats broken structural homology and also allows for the fact 
that a fixed
aperture measure for the velocity dispersion will sample different ratios of
the area enclosed by $R_{\rm e}$, as $R_{\rm e}$ itself varies between galaxies.
Here we transform the central aperture velocity dispersion measure into the 
spatial half-light velocity dispersion $\sigma _{1/2,n}$.  
% Such a treatment of the velocity dispersion term may explain 
% the differences between the observed plane and the virial plane.  
The fourth plane we construct
uses the spatial structural parameters from the $R^{1/n}$ model and an 
infinite aperture velocity dispersion term.

A Principal Component Analysis (PCA) was performed on each 
3 dimensional data set, ($\log \sigma _{0}, \log R_{\rm e,4}, \Sigma _{\rm e,4}$), 
($\log \sigma _{0}, \log R_{\rm e,n}, \Sigma _{\rm e,n}$), 
($\log \sigma _{1/2,n}, \log R_{\rm e,n}, \Sigma _{\rm e,n}$), and
($\log \sigma _{\rm tot,n}, \log R_{\rm e,n}, \Sigma _{\rm e,n}$).
The code from Murtagh \& Heck \shortcite{MaH87} was
used to show the degree to which the data are defined by a plane in our 
3-space of observables.  Table 3 shows that greater than 98 per cent of the variance
in the data lies in a plane, confirming the appropriateness of constructing
a plane to describe the properties of elliptical galaxies.

The method of construction of the plane is not an obvious task and many 
different approaches have been tried in the past \cite{DaD87,Lyn8b,LBE91}. 
In the fitting of a plane to 
3-space data, one has considerable choice in exactly what it is that one 
minimizes in order to obtain the `best' plane.  One gets a good idea of the 
problem from Isobe et al.\ \shortcite{eric1} and Feigelson \& Babu \shortcite{eric2}, where 6 possible 
approaches to the fitting of a straight line in a 2-space data set is detailed.
The situation is of course worse for higher dimensions.
However, extending the ideas learned in 2-space to 3-space, we have a better 
idea of how to approach the problem.

Our current work wishes to establish the FP for the sake of comparison with
what theory, i.e. the virial theorem, predicts.  Given this, we do
not distinguish any dependent/independent variables, and so we cannot build
our plane by minimising the residuals in one variable alone.  An approach 
that considers the residuals in all variables shall be needed, and from the
2 dimensional case studies of Isobe et al.\ \shortcite{eric1} and Feigelson 
\& Babu \shortcite{eric2}, we see that 
the bisector method of regression is most suited for our needs. For studies
requiring symmetrical treatment of the variables, the bisector method of
linear regression was found to be the preferred method, having a greater 
certainty on the slope of the fitted line than other methods, such as 
orthogonal regression.
In this technique, we construct the plane 3 times, each time minimising the
residuals of a different observable against the other two observables.  The
mean slope angles, obtained from the average of the three planes, is then used
to represent our `best' FP.

With the FP relation described as 
\begin{equation}
\log R_{\rm e} (kpc)=A\log \sigma -0.4B\Sigma _{\rm e}+C,\label{plane}
\end{equation}
the virial theorem predicts that $A$ should be 2 and $B$ should be -1.
The observed FP, as determined by the bisector method, for our four data 
sets described above, is shown in Table 4,
with the associated uncertainty being the standard error.

Because there is no natural `best' method of construction for 
the FP, we used an additional method, pioneered by Djorgovski \& 
Davis (1987), as a check.  First, the `mixing' value $b$ that gives the 
highest linear correlation 
coefficient between $\log R_{\rm e}$ and ($\log \sigma +b\Sigma _{\rm e}$) is obtained.
% pearson r was used, not Spearman's rank or Kendalls tau
%We used jackknife resampling to estimate the standard error on $b$.
% and the errors were too small ==>  .008 on b~1.250
We used the bootstrap method to estimate the standard error on $b$.
It is then possible to treat the problem as a 2-dimensional least-squares
problem.  The program SLOPES, from Feigelson \& Babu \shortcite{eric2}, was used to compute the 
best fitting line for a multitude of methods: ordinary least squares of y on
x OLS(y$\mid$x), and the inverse OLS(x$\mid$y), the line which bisects these two, 
minimisation of the perpendicular residuals (known as orthogonal regression) 
the geometric mean of OLS(y$\mid$x) and OLS(x$\mid$y) (referred to as the
reduced major axis regression line) and finally the arithmetic mean of the
OLS lines.  The reader is referred to their paper for a discussion 
of each method, for which the slopes and errors are computed using three 
techniques.
The first technique uses the linear regression formulae, but as these
can underestimate the slope variance for small data samples, a bootstrapping
simulation and a jackknife analysis are also performed.
The various solutions are displayed in Table 5.  The value of $A$ in 
equation~\ref{plane} is explicitly shown and 
the value $B$ is calculated such that $B=-2.5\times b\times A$, 
where $b$ is the `mixing' value mentioned above.

To represent the method of analysis using the mixing parameter, we use the 
orthogonal regression solution from Table 5, consistent with the method 
of Pahre et al.\ \shortcite{PDC95}, as the 
one method from the possible 6.  The bisector method should
perhaps be used, given the discussion in Feigelson \& Babu \shortcite{eric2}, 
but the differences
between these two planes are within the errors.  This orthogonal 
method gives similar planes to those found with our 3D method of analysis.
Figure~\ref{6plot} shows the various FPs constructed using our 3D bisector
method.  Figure~\ref{6plot}a uses the standard data set from the $R^{1/4}$ 
fits and the central aperture velocity dispersion measures, whilst 
Figure~\ref{6plot}b uses the data set from the $R^{1/n}$ fits and the central
aperture velocity dispersion.  Figures~\ref{6plot}c,~\ref{6plot}d use the
spatial parameters from the $R^{1/n}$ fitted model along with the 
spatial half-light velocity dispersion term and the infinite aperture velocity
dispersion term respectively.

\section{Discussion}

The additional parameter, $n$, in the $R^{1/n}$ light profile models, is shown
to have a physical association with the galaxies' other properties, such as
size and mass, as seen in Figure~\ref{enrad} and Figure~\ref{massn}.  
The bigger galaxies have larger values of $n$, and
thus less curvature in their light profiles, as previously shown by
Caon et al.\ \shortcite{CCDet} and Graham et al.\ \shortcite{Gra96}.  Not allowing for variations in profile
shape can have a substantial systematic effect on the derived half-light
radius and associated surface brightness term. Figure~\ref{radio}
shows the ratio of the half-light radius derived using the $R^{1/n}$ model to 
that obtained using the $R^{1/4}$ model, plotted against the model value
 of $n$.  One clearly sees that assuming an $R^{1/4}$ law for the larger 
galaxies
will result in an underestimate of the effective radii.  So
too, use of the $R^{1/4}$ law on the smaller galaxies will lead to
over-estimates for the galaxies' effective radii.
If the galaxy luminosity profile is best described by a value of $n$ greater 
than 4, the effective surface brightness will be too
bright if an $R^{1/4}$ profile is applied.
Likewise, the effective surface brightness term from an $R^{1/n}$ profile is 
fainter than that from the $R^{1/4}$ profile when $n$ is less than 4.
Allowing for structural differences has the result of producing a greater
range in both radius and surface brightness.

With the introduction of this additional parameter, we performed another 
Principal Component Analysis to determine the optimal number of dimensions 
in the 4-space ($n, \log R_{\rm e,n}, \Sigma _{\rm e,n}, \log \sigma _{0}$).  
The major eigenvector contained 67 per cent of the
variance, and the second dominant eigenvector had 27 per cent.  Thus, our
4-space of observables is still well represented by a two dimensional plane,
encompassing 94 per cent of the total variance (use of $\log \sigma _{1/2,n}$
and $\log \sigma _{\rm tot,n}$ gave similar results).  Therefore, a hyperplane does 
not seem warranted.  This is not surprising due to the strong correlation seen 
between $n$ and $\log R_{\rm e}$ in Figure~\ref{enrad}.

If one was to use the FP as a distance indicator, then its method of
construction would be different to the method used here.
In such a case, one would wish to predict the value of one 
distance-dependent variable from the measurement of other distance-independent 
variable(s), and so one should use OLS(y$\mid$x), where 
y is the variable to be predicted.  
With this OLS(y$\mid$x) fit we obtained an rms scatter about the
plane of 0.09 in $\log R_{\rm e}$ when an 
$R^{1/4}$ law was applied with $\sigma _{0}$, and an rms scatter of 0.10 
when an $R^{1/n}$ law and spatial structural quantities were used with 
$\sigma _{\rm tot,n}$.  The uncertainty on these
rms scatters is around 0.02, so there is no significant difference in the 
scatter about the FP using the two methods.
We also measured the intrinsic thickness of the FP by computing the 
rms orthogonal offset of the data points about the plane; here we assume no
dependent/independent variables.  These rms values are shown in Table 4.
% and suggest that the allowances for broken homology in
% the velocity dispersion profiles result in an increased scatter about the
% plane.  We again note the above caution on the size of the increase.
% We computed the rms scatter of the data points, orthogonal to the plane.
% The intrinsic width of the FP, is smaller than that given above (because
% of measurement errors and a finite cluster size, which will increase
% the scatter -- in quadrature)   See my email to George Djorgovski.

%Both our 3 dimensional bisector method and the mixing parameter approach 
%of Djorgovski produce similar results for the FP.

The virial theorem predicts the relation $R\propto \sigma ^{2}\Sigma ^{-1}$, when
the surface brightness, $\Sigma $, is expressed in linear units rather than mag 
arcsec$^{-2}$.  Observed planes for cluster galaxy data, using $R=R_{\rm e}, 
\Sigma =\Sigma _{\rm e}$, and 
$\sigma =\sigma _{0}$ (the projected central velocity dispersion) have been 
in good agreement with each other in the past.
Dressler et al.\ (1987b) found $R\propto \sigma ^{1.33}\Sigma ^{-0.83}$ for 6 combined
galaxy clusters, with Lucey et al.\ \shortcite{LBE91} obtaining 
$R\propto \sigma ^{1.27}\Sigma ^{-0.78}$ for a Virgo subsample ($\mu _{\rm e}$(B)$<$22.00) 
of 16 ellipticals.  Lucey et al.\ \shortcite{LGCT9} found a similar result for their $V$-band study 
of 51 ellipticals in Coma, with $R\propto \sigma ^{1.23}\Sigma ^{-0.82}$ and a 
larger study of 11 clusters by J\o rgensen, Franx \& Kj\ae rgaard 
\shortcite{Jor96}, taken in Gunn {\it r}, produced an identical result.  

A recent near-infrared study of 5 nearby clusters (59 elliptical galaxies) 
by Pahre et al.\ \shortcite{PDC95} gave $R\propto \sigma ^{1.44}\Sigma ^{-0.79}$. 
This steeper 
slope of the plane was attributed to a reduction of metallicity effects in
the near-infrared bandpass compared to the optical.
Intriguingly, they obtained the notably different 
$R\propto \sigma ^{1.62}\Sigma 
^{-0.93}$ for their Virgo sample of galaxies.  As this sample 
only numbered 8 galaxies, this result does not carry much weight on its own,
but does warrant further investigation.  

As shown in Table 4, allowing for structural homology through use of an 
$R^{1/n}$ profile, and maintaining use of the central velocity dispersion, 
{\it actually increases} the departure of the FP from the virial plane.
Whilst the radial term changes in the sense that it reduces the departure, the
surface brightness term works to increase the departure and proves to be the
dominant effect.  However, changes to the structure of a galaxy imply changes
to the velocity field of the galaxy.  These changes can be obtained through 
application of the Jeans equation for a pressure-supported system.

One problem with the use of central velocity dispersions
is that they sample different portions of the half-light radius for 
each galaxy.  To be consistent in our measuring of galaxy parameters, 
we computed all parameters at the spatial half-light radius.  This meant 
converting central aperture velocity dispersions into the associated value
at the spatial half-light radius (see Appendix B for details).   Doing this,
and using the spatial parameters from the $R^{1/n}$ profiles, we found that the 
FP constructed from these quantities more closely resembled the virial plane.
The exponent on the surface brightness term in this expression for the FP is 
consistent with the virial theorem prediction of -1.  Past differences
between this exponent and the virial expectation can be explained by systematic 
departures from structural and dynamical homology and the use of projected 
rather than volumetric velocity dispersion.  The exponent on the $\sigma $ term 
is closer to the value expected from the virial theorem, changing from 1.33 to
1.48 as we allow for such changes, but it does not reach the expected value 2.

Although use of the spatial half-light velocity dispersion for all galaxies 
is consistent 
and methodical, and preferred to the use of central velocity dispersion
measures, it is still arbitrary.  We have therefore derived the total aperture
velocity dispersion (i.e.\ the value of the aperture velocity dispersion at 
infinite radius), which for spherical systems is equal to one-third of the 
(three dimensional) virial velocity dispersion (Ciotti 1994).
%Our galaxy data set, taken from 
%Bower et al.\ (1992), was imaged in the V-band, so we, and most studies 
%before us, will be somewhat 
%more contaminated by possible metallicity effects than an IR study.
%However, as we construct the FP using both $R^{1/4}$ and $R^{1/n}$ profiles,
%the change in the FP due to structural effects will still be apparent.
The FP that allows for differences in galaxy profile shape and uses the total 
aperture velocity dispersion more closely resembles the virial plane, 
$R\propto \sigma ^{2}\Sigma ^{-1}$, than the standard observed plane 
obtained with $R^{1/4}$ law parameters and central aperture velocity 
dispersions.  The respective planes are described by 
$R\propto \sigma ^{1.44\pm0.11}\Sigma ^{-0.93\pm0.08}$ and
$R\propto \sigma ^{1.33\pm0.10}\Sigma ^{-0.79\pm0.11}$.

Capelato et al.\ (1995) found, using computer simulations to build and merge
elliptical galaxies, that as they increased the aperture size in which they 
measured the velocity dispersion term, the difference between the FP they built
and that predicted by the virial theorem was reduced.  Although the initial 
conditions of their models were some-what limited and their range of merged 
galaxies does not fully represent the range of elliptical galaxies we see 
today, their work strongly indicates that the assumption of homology amongst 
elliptical galaxies is invalid.  They effectively held constant the stellar 
population, did not alter 
the IMF or M/L ratios, and have no metallicity effects to worry about in their 
dissipationless models.  The tilt of the FP in their models is therefore not due
to any of these effects.
%Consequently, one can expect use of aperture sizes that 
%are different ratios of $R_{\rm e}$, which are used to measure central velocity 
%dispersions, will have an influence on the derived FP. Our use of a 
% This would also be the case even if all galaxies were $R^{1/4}$.
Their work implied that most of the nonhomology lies close to the 
center of their merged galaxy models, and hence use of larger apertures 
gives a more
accurate picture of the virialised system --- thus the reduction between
the observed and virialised planes found with the use of a larger aperture.  
A similar physical effect could be 
occurring with our use of an effectively larger aperture, albeit 
an infinite aperture dispersion based on transforming
the central aperture dispersion using the $R^{1/n}$ light models, 
compared to the average aperture size
of only 0.2$R_{\rm e}$ that has been used in the past.  

Although our model 
velocity dispersion profiles allow for different distributions,
as described by the $R^{1/n}$ model, they are built upon certain assumptions, 
and we have not dealt with possible orbital anisotropies, as would have
been done implicitly in the computational work of Capelato et al.\ (1995).  
In examining the structural and dynamical deviations from homology between the 
progenitor and merged galaxy models, Capelato et al.\ (1995) claim 
the nonhomologous velocity dispersion profiles as the main cause of the tilt 
to the FP.  However, their merger remnants only spanned a range in $n$ from 1.2 to 
3.4, so they have not really explored the full range of structures 
observed in nature, and cannot entirely exclude this as a contributing factor
to the tilt of the FP.

Ciotti et al.\ \shortcite{CLR96} and Ciotti \& Lanzoni \shortcite{CaL96} 
concluded that velocity anisotropy cannot
cause the observed FP tilt by itself (i.e. if structural homology is assumed).
This result was derived from a range of theoretical galaxy models (including 
the $R^{1/n}$ model) and introducing varying amounts of orbital anisotropy as 
described by the Osipkov-Merritt parameterisation \cite{Osi79,Mer85}.  
Ciotti \& Lanzoni \shortcite{CaL96} did 
however suggest an observational test for possible velocity anisotropy effects,
namely the use of large apertures to measure the velocity dispersion term,
an idea supported by Capelato et al.\ (1995) and incorporated in this study.
Our work suggests that the combination of broken homology in both 
the surface brightness and dispersion profiles of elliptical systems is partly 
responsible for the departure of the observed FP from the predicted plane.

%Obtaining velocity dispersion profiles is a difficult task for the 
%observational astronomer, and certain assumptions, such as in our work, have 
%been necessary to date.
%\cite{PaD96} hope to use measured velocity dispersion profiles, investigating 
%both the nonhomologous nature of elliptical galaxies and possible 
%aperture effects as described above.

These results have implications for most studies that have been done to date
using the FP, or the $D-\sigma $ relation, because such studies have assumed 
structural homology and used projected quantities.
Studies of the peculiar motions of elliptical galaxies in the local Universe 
\cite{Lyn8b,LaC88,DaF90} now warrant re-examination, as
such measurements may be systematically affected by biases in the distance
indicator relations.
Furthermore, studies of environmental dependence on the $D-\sigma $ relation 
and differences between field and cluster ellipticals 
\cite{Sil89,DDH88,LGCT9,BFD90,CaD92} can be re-examined.  
If the galaxy structure 
is dependent on the galaxy environment, then the $D-\sigma $ relation may 
also be dependent on environment.  However, in fitting an $R^{1/n}$ profile, 
we have already taken into account different galaxy structures (and so possibly
different galaxy environments).  Thus the new
FP which we derive may implicitly take into consideration some 
environmental effects, although not stellar population differences 
(Guzman \& Lucey 1993; Guzman et al.\ 1993) or 
intrinsically different $M/L$ ratios in different clusters \cite{Kai88}.
It will of course be of interest to obtain data from other clusters, and field
galaxies, to investigate such possibilities.

Working with the simplification that the $K_{SR}$ term in 
equation~\ref{kvirl} is now a constant 
(although it still contains the $k_{\rm e}$ and $k_{V}$ terms which we have not 
properly dealt with) we can compute the mass-to-light ratio, assuming it is a 
power law
function of the observables.  For the plane constructed from the $R^{1/4}$
law model and using the central aperture velocity dispersion measure, we 
have that $M/L\sim L^{0.25}\Sigma ^{-0.06}$; whereas for the 
plane that was constructed allowing for broken structural homology and using
the volumetric quantities, we find
that $M/L\sim L^{0.17}\Sigma ^{0.17}$, or that 
$M/L\sim \sigma^{1/2}$.  For the plane that used $R^{1/n}$ model parameters
and the total aperture velocity dispersion measure we have that 
$M/L\sim L^{0.20}\Sigma ^{0.09}$. 
Of course it may be that M/L is constant and it
is the $K_{SR}$ term in equation~\ref{kvirl} that varies with $L$ and $\Sigma $ 
or $\sigma $, as mentioned previously, or it may be a that terms are varying.  

The tilt of the FP is probably due to a combination of factors.
For instance, whilst Pahre et al.\ \shortcite{PDC95} did not recover the virial 
plane in their
construction of the near-infrared FP, they did measure a change in $\alpha $, 
with $M/L\sim M^{\alpha }$,
from 0.23 \cite{LGCT9,CaD92} in the optical (V-band) to $\alpha $=0.16 
(K-band).  They found that the exponent on the velocity dispersion term, A,
increased by $0.19\pm 0.06$ dex from the V-band to the K-band 
(cf. $+0.29\pm 0.11$, Djorgovski \& Santiago  \shortcite{DaS93}).  
%It is expected
%that our result might also change by a similar amount, once we have obtained
%near-infrared data; thus further reducing the departure between the observed
%and the virial plane.

Whilst we have dealt with the effects of seeing on the galaxy light profile, by
excluding data within the central radius of 3 FWHM, we have not dealt
with the seeing effect on the central velocity dispersion measurements.  As an
effective 16$\times$16 arcsec aperture was used to obtain the velocity
dispersion for our sample (Dressler 1984, 1987), 
seeing should not be a problem.  However,
studies using a single smaller aperture (say 2$\times$4 arcsec)
may be effected by seeing if there is a steep velocity dispersion gradient
present \cite{Whi80,JaP91}.

Several other issues must be kept in mind: the galaxies may not be
virialised; we have not dealt with the matter of rotational support
(as opposed to pressure support);
the issue of broken dynamical homology has not been fully treated, as we 
assumed orbital isotropy; 
% Capelato et al.\ (1995) have shown through computer simulations
% of merging galaxies, that a non-homologous velocity distribution is an
% important effect in causing deviations of the observed plane from the virial
% plane.  However, Ciotti et al.\ \shortcite{CLR96} conclude that dynamical 
% anisotropy variations along the FP cannot alone explain the deviation.  
% Ciotti \& Lanzoni \shortcite{CaL96} have investigated dynamical anisotropy 
% applied to the $R^{1/n}$ galaxy light profiles and confirm their previous 
% result.
stellar population variations with galactic radius may result in (M/L)
varying with radius; or light may not trace mass in a simple 
fashion. Any combination of these effects may work in symphony to account 
for the FP tilt.

\section{Conclusions}

A range in profile shapes is found to exist for the sample of 26 elliptical 
galaxies used in this study.  The shape parameter $n$ is found to have a 
physical association with galaxy size and mass, in the sense that a larger value
of $n$ corresponds to a larger galaxy.  Allowances for 
different profile shapes and the use of infinite aperture velocity dispersion
measurements, which allow for different velocity dispersion profile shapes, 
results in a modification to the FP.
While more clusters need be studied, our work shows that broken structural
homology is not solely responsible for the tilt of the FP, but allowing 
for variations caused by such structural differences does reduce the tilt.
It is likely that the departure of the observed FP from the virial plane 
will be further reduced by using infrared 
luminosity profile measurements to minimise stellar population effects.
Our best estimate of the FP using the parameters from an $R^{1/n}$ profile and
the total aperture velocity dispersion is 
$R\propto \sigma ^{1.44\pm0.11}\Sigma ^{-0.93\pm0.08}$.  The exponent of the
surface brightness term is consistent with the prediction of the virial theorem. 
% where $R, \sigma $ 
% and $\Sigma $ are spatial parameters from $R^{1/n}$ profile models.  
% Assuming equal degrees of virialisation for the galaxies in our sample, 
% and that dynamical homology is present throughout, this implies 
% $M/L\sim \sigma^{1/2}$.  
% However, it is unlikely that these conditions apply, and so we do not place 
% too much emphasis on the above relationship.

\subsection*{Acknowledgments}

We thank Richard Bower, John Lucey, and Richard Ellis for kindly making their 
galaxy profile data available to us in electronic form.
We are grateful for discussions with George Djorgovski, and 
Reinaldo de Carvalho.  We also appreciate the useful suggestions and comments
from Mike Pahre that helped to improve this paper.
We are grateful for Luca Ciotti's suggestion too explore use of the total 
aperture velocity dispersion in our study.
We wish to thank Eric Feigelson and Michael Akritas at the 
Statistical Consulting Centre for Astronomy at Pennsylvania State University,
for their advice on the construction of the Fundamental Plane.
We are grateful for the use of Eric Feigelson's computer code SLOPES.
This research has made use of the NASA/IPAC Extragalactic Database (NED)   
which is operated by the Jet Propulsion Laboratory, California Institute   
of Technology, under contract with the National Aeronautics and Space      
Administration.

\clearpage
\begin{figure}
\caption{The residual V-band aperture magnitude data (beyond 3 FWHM, and taken from Bower, Lucey \& Ellis (1992)) about the best fitting $R^{1/4}$ and $R^{1/n}$ profile models is shown, being open triangles and solid squares respectively.\label{curve}}
\caption{The average shape parameter $n$, from the $R^{1/n}$ models, is plotted against the average spatial half-mass radius for the 26 E/S0 Virgo galaxies from Bower, Lucey, \& Ellis (1992).\label{enrad}} 
\caption{The shape parameter $n$, from the $R^{1/n}$ models, is plotted against the projected effective radius for all the the images (including multiples).  Also shown are the 1$\sigma $ error ellipses, as described in the text.\label{error}}
\caption{The $\chi ^{2}$ value of the best fitting $R^{1/n}$ (filled square), and ($R^{1/4} +$ sky-correction) (star), model, normalised against the $\chi ^{2}$ value of the best fitting $R^{1/4}$ model, is plotted against $n$.  The ratio is of course close to 1 near $n=4$, and is smaller when away from $n$=4, indicating a superior fit of the model to the data. \label{chise}}
\caption{The $\chi ^{2}$ value of the best fitting ($R^{1/4} +$ sky-correction) model, normalised against the $\chi ^{2}$ value of the best fitting $R^{1/4}$ model, is plotted against the size of the sky correction expressed as the associated change in the half-light magnitude.\label{ratts}}
\caption{The ratio of the $\chi ^{2}$ fit from the best fitting $R^{1/n}$ model to the best fitting $R^{1/4}$ model with a sky-correction, is plotted against the size of the sky correction expressed as the associated change in the half-light magnitude.\label{ratio}}
\caption{The spatial ($\sigma (r)$, solid), line-of-sight ($\sigma _{ls}(R)$, dashed), and aperture ($\sigma _{\rm ap}(R)$, dotted) velocity dispersion profiles are shown for a range of shape parameters $n$ for the $R^{1/n}$ luminosity profile. The abscissa is such that $s=r/R_{\rm e}$ and $\eta =R/R_{\rm e}$. \label{vel}}
\caption{The ratio of the aperture velocity dispersion, as calculated in Appendix B, to the spatial half-mass velocity dispersion, as a function of projected aperture size. The abscissa is such that $\eta =R/R_{\rm e}$\label{perc}}
\caption{Total galaxy mass plotted against the shape parameter $n$.\label{massn}}
\caption{a) The FP constructed using an $R^{1/4}$ law to derive the spatial half-mass radius (R) and mean surface brightness ($\mu $); and using the central projected aperture velocity dispersion measure for $\sigma $.  b) Similar to (a) except that $R^{1/n}$ profile parameters were used.  c) Similar to (b) except that the spatial half-light velocity dispersion term is used instead of the central aperture velocity dispersion.  d) Constructed using the parameters from an $R^{1/n}$ law and the total (infinite) aperture velocity dispersion term.\label{6plot}} 
\caption{Ratio of the derived effective radii from the $R^{1/n}$ and the $R^{1/4}$ model is shown against the shape parameter $n$.  A clear systematic effect of under and over estimation of the effective radius is evident when applying the $R^{1/4}$ law to large and small galaxies respectively.\label{radio}}
\end{figure}

\clearpage
\appendix
\section{Derivation of $\langle\mu\rangle_{\rm e}$ from an $R^{1/n}$ curve of growth}

From equation~\ref{sereq} the Sersic law \cite{Ser68}, describing the 
projected light distribution for elliptical galaxies, bulges and disks is 
given by
\begin{equation}
I=I_{\rm e}e^{b}e^{-b\left(R/R_{\rm e}\right)^{1/n}}.
\end{equation}
The average intensity, $<$$I$$>$$_{\rm e}$, within the effective radius, 
$R_{\rm e}$, of the model is obtained from
\begin{equation}
<I>_{\rm e}=\frac{\int I dA}{A}=\frac{I_{\rm e}e^{b}\int _{0}^{R_{\rm e}}e^{-b\left(R/R_{\rm e}\right)^{1/n}} 2\pi R dR}{\pi R_{\rm e}^{2}}.
\end{equation}
This expression can be simplified by letting $x=b(R/R_{\rm e})^{1/n}$, giving
\begin{equation}
<I>_{\rm e}=I_{\rm e}f(n),
\end{equation}
where
\begin{equation}
f(n)=\frac{2n e^{b}}{b^{2n}}\int_{0}^{b}e^{-x}x^{2n-1} dx.
\end{equation}
Now as $b$ was chosen such that $R_{\rm e}$ is the radius containing
half of the total light from the galaxy, we have
\begin{equation}
f(n)=\frac{n e^{b}}{b^{2n}}\int_{0}^{\infty}e^{-x}x^{2n-1} dx=\frac{n e^{b}}{b^{2n}}\Gamma (2n)=\frac{k_{L}}{\pi },
\end{equation}
where $k_{L}$ is the luminosity structure term mentioned in section~\ref{3two}.
By taking the logarithm, we obtain
\begin{equation}
\langle\mu\rangle_{\rm e}=\mu _{\rm e}-2.5\log [f(n)].
\end{equation}
The value of $\mu _{\rm e}$ comes straight from the fitted $R^{1/n}$ model, and 
$f(n)$ is computable and shown in Figure~\ref{appA1} for a range of $n$ that 
covers typical observational data.  This result is in agreement with the
previous work of Caon et al.\ \shortcite{Cao94}.

\begin{figure}
\caption{The structural luminosity term $K_{L}$ (s.t. -2.5$\log(K_{L})=\langle\mu\rangle_{\rm e}-\mu _{\rm e}-2.5\log(\pi )$) is plotted as a function of $n$, for Sersic's $R^{1/n}$ law.  The dashed line aids the eye for the case where $n=4$, giving the result for the de Vaucouleurs $R^{1/4}$ law.\label{appA1}} 
\end{figure}

\clearpage
\section{Velocity Dispersion Profiles}

The spatial luminosity density $\nu (r)$ can be computed from the projected 
luminosity profile $I(R)$ of the galaxy, such that
\begin{equation}
\nu (r)=\frac{-1}{\pi}\int _{r}^{\infty}\frac{dI(R)}{dR}\frac{dR}{\sqrt{R^{2}-r^{2}}} ,
\end{equation}
where R is the projected radius and r is the spatial radius of the galaxy
\cite{BaT87}. 

This has been done for the $R^{1/4}$ profile \cite{Pov60,You76},
and the generalisation to the $R^{1/n}$ profile (Ciotti 1991) is
\begin{equation}
\nu _{n}(s)=\frac{b^{n}x^{1-n}}{\pi }\int _{0}^{1}\frac{1}{t^{2}}\frac{\exp(-x/t)}{\sqrt{t^{-2n}-1}}dt,
\end{equation}
where $s=r/R_{\rm e}$ and $x=bs^{1/n}$.  We have replaced the dummy variable $t$ 
in Ciotti's expression by $1/t$, so as to avoid an integration to infinity.  
The physical value for the luminosity density is given by 
$\nu (r)=I_{\rm e}e^{b}/R_{\rm e}\, \nu (s)$.  Following 
Ciotti's Figure 2, Figure~\ref{appB1} shows the luminosity density as a 
function of the normalised radius $s$, for a range of shape parameters $n$.  

The integrated spatial luminosity is then
\begin{equation}
L_{n}(s)=4\pi \int _{0}^{s}\nu _{n}(s')s'^{2}ds',
\end{equation}
and is shown in Figure~\ref{appB2}, with $n$ ranging from 1 to 15.
Here we have that $L_{n}(r)=I_{\rm e}e^{b}R_{\rm e}^{2}\, L_{n}(s)$, where 
$L_{n}(r)$ yields the same total luminosity as given by equation~\ref{ltot}. 
We include Figures B1 and B2 in our paper for
ease of reference and as they cover a greater range of shape parameters than in
Ciotti's Figure 2 ($n=2-10$), where we 
encompass the $n=1$ model being representative of an exponential disk and all 
integer values up to $n=15$ being the cut off used by Caon et al.\ (1993).

For a spherical galaxy, with isotropic dynamics, constant M/L, and in 
hydrostatic equilibrium, the Jeans hydrodynamical equation can be written as
$d(\nu \sigma ^{2})/dr=G(M/L)L(r)\nu /r^{2}$.
Integration with the boundary condition $\nu \sigma ^{2}\rightarrow 0$ as 
$r\rightarrow \infty $ gives
\begin{equation}
\sigma _{n}^{2}(s)=\frac{1}{\nu _{n}(s)}\int _{0}^{1/s}L_{n}(1/t)\nu _{n}(1/t)dt,
\end{equation}
where we have dropped the constants $G(M/L)$.  Again, we have replaced the 
dummy variable $t$ with $1/t$.  This expression for the velocity dispersion
 is dimensionless, the physical value $\sigma _{n}^{2}(r)$ being given by
\begin{equation}
\sigma _{n}^{2}(r)=G(M/L)I_{\rm e}e^{b}R_{\rm e}\, \sigma _{n}^{2}(s).
\end{equation}

The line of sight velocity dispersion is given by Ciotti \shortcite{Cio91} as 
\begin{equation}
\sigma _{ls, {\it n}}^{2}(\eta )=-2e^{b\eta ^{1/n}}\int _{0}^{1/\eta }L_{n}(1/t)\nu _{n}(1/t)\sqrt{1/t^{2}-\eta ^{2}} dt,
\end{equation}
where $\eta =R/R_{\rm e}$.
The luminosity weighted aperture velocity dispersion is then
\begin{equation}
\sigma _{ap, {\it n }}^{2}(\eta _{\rm ap})=\frac{\int _{0}^{\eta _{\rm ap}}I(\eta )\sigma _{ls}(\eta )\eta d\eta }{\int _{0}^{\eta _{\rm ap}}I(\eta )\eta d\eta },
\end{equation}
where $I(\eta )$ is the $R^{1/n}$ profile.
The physical quantity $\sigma _{ap, {\it n }}^{2}(R_{\rm ap})$ being given
by the same transformation, such that
\begin{equation}
\sigma _{ap, {\it n }}^{2}(R_{\rm ap})=G(M/L)I_{\rm e}e^{b}R_{\rm e}\, \sigma _{ap, {\it n }}^{2}(\eta _{\rm ap}).\label{maseq}
\end{equation}

Shown in Figure~\ref{vel} are the spatial, line of sight, and aperture 
velocity 
dispersion as a function of radius for different $R^{1/n}$ profiles.  It 
is noted that this quantity is dimensionless and should be multiplied by 
$G(M/L)I_{\rm e}e^{b}R_{\rm e}$ to obtain the physical quantity.
The ratio of the spatial velocity dispersion at the spatial half-mass radius, 
$s\sim 1.34$, to the aperture velocity
dispersion at the value of $\eta $ corresponding to the ratio of the 
aperture size used and and the projected half-light radius of the galaxy, 
can be used to correct the measured central velocity dispersion to the 
physical value at the spatial half-mass radius.  Thus, the multiplicative 
factor doesn't come into play in this conversion process.

\begin{figure}
\caption{The spatial luminosity density ($\nu $) profiles are shown for a range in $n$ from the exponential model, where $n=1$, to $n=15$, with $s=r/R_{\rm e}$ being the 3D spatial radius divided by the projected half-light radius.  (Extension of Fig.\ 2 from Ciotti (1991) which plotted values of $n$ ranging from 2 to 10).\label{appB1}}
\caption{The integrated spatial luminosity, normalised by the total profile luminosity, is plotted as a function of $s=r/R_{\rm e}$. (Extension of Fig.\ 2 from Ciotti (1991) which plotted values of $n$ ranging from 2 to 10).\label{appB2}}
\end{figure}

\end{document}